\documentclass[a4paper,english,aps,preprint,nofootinbib,superscriptaddress]{revtex4}
\usepackage[T1]{fontenc}
\usepackage[latin9]{inputenc}
\usepackage{xcolor}
\usepackage{pdfcolmk}
\usepackage{babel}
\usepackage{array}
\usepackage{multirow}
\usepackage{amsmath}
\usepackage{amssymb}
\usepackage{graphicx}
\PassOptionsToPackage{normalem}{ulem}
\usepackage{ulem}
\usepackage[unicode=true,
 bookmarks=true,bookmarksnumbered=true,bookmarksopen=true,bookmarksopenlevel=1,
 breaklinks=false,pdfborder={0 0 0},backref=false,colorlinks=true]
 {hyperref}

\makeatletter

\pdfpageheight\paperheight
\pdfpagewidth\paperwidth

\providecommand{\tabularnewline}{\\}
\providecolor{lyxadded}{rgb}{0,0,1}
\providecolor{lyxdeleted}{rgb}{1,0,0}
\@ifundefined{textcolor}{}
{%
 \definecolor{BLACK}{gray}{0}
 \definecolor{WHITE}{gray}{1}
 \definecolor{RED}{rgb}{1,0,0}
 \definecolor{GREEN}{rgb}{0,1,0}
 \definecolor{BLUE}{rgb}{0,0,1}
 \definecolor{CYAN}{cmyk}{1,0,0,0}
 \definecolor{MAGENTA}{cmyk}{0,1,0,0}
 \definecolor{YELLOW}{cmyk}{0,0,1,0}
}

\usepackage{bbm}
\usepackage{bm}
\usepackage{slashed}

\makeatother

\begin{document}

\preprint{FTUV-18-0615.1693, IFIC/18-22}

\title{Fitting flavour symmetries: \\
the case of two-zero neutrino mass textures}

\author{Julien Alcaide}

\email{julien.alcaide@uv.es}

\affiliation{Departament de Física Teòrica, Universitat de València\\
 and IFIC, Universitat de València-CSIC\\
Dr. Moliner 50, E-46100 Burjassot (València), Spain}

\author{Jordi Salvado}

\email{jsalvado@icc.ub.edu}

\affiliation{Departament de Física Quàntica i Astrofísica and Institut de Ciencies
del Cosmos, \\
Universitat de Barcelona, \\
Diagonal 647, E-08028 Barcelona, Spain}

\author{Arcadi Santamaria}

\email{arcadi.santamaria@uv.es }

\affiliation{Departament de Física Teòrica, Universitat de València\\
 and IFIC, Universitat de València-CSIC\\
Dr. Moliner 50, E-46100 Burjassot (València), Spain}
\begin{abstract}
We present a numeric method for the analysis of the fermion mass matrices
predicted in flavour models. The method does not require any previous
algebraic work, it offers a $\chi^{2}$ comparison test and an easy
estimate of confidence intervals. It can also be used to study the
stability of the results when the predictions are disturbed by small
perturbations. We have applied the method to the case of two-zero
neutrino mass textures using the latest available fits on neutrino
oscillations, derived the available parameter space for each texture
and compared them. Textures $A_{1}$ and $A_{2}$ seem favoured because
they give a small $\chi^{2}$, allow for large regions in parameter
space and give neutrino masses compatible with Cosmology limits. The
other ``allowed'' textures remain allowed although with a very constrained
parameter space, which, in some cases, could be in conflict with Cosmology.
We have also revisited the ``forbidden'' textures and studied the
stability of the results when the texture zeroes are not exact. Most
of the forbidden textures remain forbidden, but textures $F_{1}$
and $F_{3}$ are particularly sensitive to small perturbations and
could become allowed.

\end{abstract}
\maketitle

\section{Introduction}

Understanding fermion masses and mixings is probably one of the most
stubborn problems the particle physics community has nowadays: we
have plenty of data about masses and mixings, which present clear
patterns of hierarchies, yet we are unable to understand their origin
and their values. The Standard Model (SM) just parametrizes them with
complete generality and satisfying all requirements of renormalizable
quantum field theories. The most popular theories beyond the SM (supersymmetry
for instance) do not add much on the subject. The solution of this
problem is probably linked to the origin of the spontaneous symmetry
breaking mechanism in the SM or to the question on why there are only
three generations of fermions. Until the complete solution is found
one may adopt a more modest bottom-up approach and try to find patterns
that relate the many parameters that characterize flavour. One of
the simplest approaches in this direction has been to find texture
zeroes in the mass matrices that are compatible with the data (see
\cite{Wilczek:1977uh,Fritzsch:1977za,Ramond:1993kv} for the quark/lepton
sector and \cite{Frampton:2002yf} for the neutrino sector). These
texture zeroes are supposed to be enforced by a symmetry (see for
instance \cite{Grimus:2004hf,Dev:2011jc}) or be approximate statements
dictated by the dynamics of a more complete theory (for instance in
many radiative neutrino mass models \cite{Zee:1985id,Babu:1988ki}
neutrino masses can be computed and are proportional to the charged
lepton masses, in that case the elements proportional to $m_{e}$
are expected to be much smaller than the others, see for instance
\cite{delAguila:2011gr,delAguila:2012nu,Alcaide:2017xoe}). Here we
will discuss Majorana neutrino mass textures in the spirit of \cite{Frampton:2002yf}
in which one looks for zeroes of the neutrino mass matrix in a basis
in which the charged lepton mass matrix is already diagonal (the analysis
of the texture zeroes with arbitrary charged lepton and neutrino mass
matrices is much more complicated for it can be shown that some textures
are trivial in the sense that they can be obtained from general matrices
just by changing the flavour basis \cite{Branco:2007nn}). In particular,
two-zero textures are very interesting because they give four relations
among the nine real parameters needed to describe the Majorana neutrino
mass matrix, and these relations can be checked against available
data. In ref.~\cite{Frampton:2002yf} it was shown that there are
only seven two-zero textures which can accommodate data on neutrino
masses and mixings (all three-zero textures were already excluded).
These textures have been extensively studied in the past (see \cite{Guo:2002ei,Dev:2006qe,Fritzsch:2011qv,Meloni:2012sx,Kitabayashi:2015jdj,Zhou:2015qua,Singh:2016qcf}
for recent analyses). 

In most of the works the relations among parameters have been derived
analytically for the different textures. These relations have been
used to scan the parameter space, letting the six parameters measured
in neutrino oscillation experiments vary in their allowed $3\sigma$
regions and checking if the two-zero texture relations are satisfied.
In general, correlations among oscillation parameters are neglected.
This is a good approximation for most of them, but we now know the
exact shape of the allowed region in the parameters $s_{23}^{2}$\textendash $\delta$
($s_{23}^{2}=\sin^{2}\theta_{23}$ is the $2-3$ mixing, and $\delta$
the Dirac phase in the neutrino mixing matrix) is quite asymmetric.
This is in part due to the octant ambiguity in $s_{23}^{2}$ and the
asymmetry in $\delta$ due to matter effects. 

Here we will present an extremely simple method, completely numerical
since the beginning, and will use previous results as a testbed for
the method. The method, based on the minimization of a generalized
$\chi^{2}$ function, which incorporates the constraints imposed by
the textures, is now possible thanks to the fact that the NuFIT collaboration
\cite{Esteban:2016qun,nufitweb}~\footnote{See also \cite{deSalas:2017kay,Capozzi:2018ubv} for alternative recent
fits to neutrino oscillation data.} has made publicly available the $\Delta\chi^{2}$ of their fits to
neutrino oscillation data, and to the new Monte Carlo tools as MultiNest
\cite{Feroz:2008xx,Feroz:2013hea} that will allow us a very robust
and efficient scanning of the parameter space. The method incorporates
naturally $s_{23}^{2}$\textendash $\delta$ correlations, allows
us to compute the available parameter space after the texture zeros
have been imposed and provides a standard $\chi^{2}$ comparative
test of how well the different textures can accommodate the experimental
data~\footnote{For a related approach based also on a $\chi^{2}$ analysis see \cite{Meloni:2012sx}.}.
It also generalizes trivially to the case in which the zeroes are
only approximate \footnote{One expects that, in some cases, radiative corrections will shift
the texture zeroes to some small quantities~\cite{Hagedorn:2004ba}.}. All this without the need of any previous algebraic work to disentangle
the relations among parameters. 

Although we have concentrated in neutrino mass two-zero textures using
the last oscillations data available (NuFIT 2018, release 3.2 \cite{nufitweb}),
an aim of this work is to provide a general template to analyze numerically
the different flavour models which give predictions for neutrino masses.
Therefore, the methods developed here can be applied to other neutrino
mass models and also to quark mass matrices with texture zeroes or
other constraints.

Thus, in section \ref{sec:The-two-zero-textures} we fix the notation
and briefly introduce the different two-zero neutrino mass matrix
textures. In section~\ref{sec:The-method} we present the method
we use to analyze the textures while in section \ref{sec:AllowedTextures}
we study the available parameter space for all the allowed textures.
In section~\ref{sec:Approximate-texture-zeroes} we discuss how the
results change if the texture zeroes are only approximate and also
review non-allowed textures when the zeroes are only approximate.
Finally in section \ref{sec:Conclusions} we collect the main conclusion
of our analysis.

\section{The two-zero textures\label{sec:The-two-zero-textures}}

Two-zero neutrino textures in the \cite{Frampton:2002yf} approach
are defined in the basis in which the charged lepton Yukawa matrices
are diagonal and there are only three active neutrino characterized
by a Majorana neutrino complex symmetric matrix \footnote{Generalization to texture zeroes including sterile neutrinos\cite{Krolikowski:1999cx,Zhang:2013mb,Nath:2015emg},
zeroes of the inverted neutrino mass matrix \cite{Lavoura:2004tu},
or other relations among matrix elements \cite{Branco:2002ie,He:2003nt,Kaneko:2005yz,Chauhan:2006uf,Lashin:2007dm,Han:2017wnk}
is possible.}. In this basis, the neutrino mass matrix can be reconstructed from
the six neutrino oscillation parameters, $s_{12}^{2}$, $s_{23}^{2}$,
$s_{13}^{2}$, $\Delta_{21}=m_{2}^{2}-m_{1}^{2}$, $\Delta_{31}=m_{3}^{2}-m_{1}^{2}$
and $\delta$, and three more parameters still unknown, the lightest
neutrino mass $m_{\ell}$ (which is equal to $m_{1}$ in the normal
ordering (NO) solution and $m_{3}$ in the inverted one (IO)\footnote{We use the conventions of \cite{Esteban:2016qun,nufitweb} in which
$\Delta_{31}$ is replaced by $\Delta_{32}$ $=m_{3}^{2}-m_{2}^{2}$
in the IO case.}) and two Majorana phases, $\alpha_{1}$ and $\alpha_{2}$:
\begin{equation}
M_{\nu}\equiv\begin{pmatrix}M_{ee} & M_{e\mu} & M_{e\tau}\\
M_{e\mu} & M_{\mu\mu} & M_{\mu\tau}\\
M_{e\tau} & M_{\mu\tau} & M_{\tau\tau}
\end{pmatrix}=UD_{\nu}U^{T}\;,\quad{\rm with}\;\;D_{\nu}=\mathrm{diag}(m_{1},m_{2},m_{3})\label{eq:masses}
\end{equation}
where $U$ is the PMNS matrix and can be written as 
\begin{eqnarray}
U=\left(\begin{array}{ccc}
c_{13}c_{12} & c_{13}s_{12} & s_{13}e^{-i\delta}\\
-c_{23}s_{12}-s_{23}s_{13}c_{12}e^{i\delta} & c_{23}c_{12}-s_{23}s_{13}s_{12}e^{i\delta} & s_{23}c_{13}\\
s_{23}s_{12}-c_{23}s_{13}c_{12}e^{i\delta} & -s_{23}c_{12}-c_{23}s_{13}s_{12}e^{i\delta} & c_{23}c_{13}
\end{array}\right)\left(\begin{array}{ccc}
e^{i\alpha_{1}/2}\\
 & e^{i\alpha_{2}/2}\\
 &  & 1
\end{array}\right)\ ,\label{eq:UPMNS}
\end{eqnarray}

Two-zero textures impose that $(M_{\nu})_{ab}=0$ for two different
elements. There are ${\displaystyle {\displaystyle \binom{6}{2}=15}}$
two-zero neutrino textures which were classified in ref. \cite{Frampton:2002yf}
in two groups; allowed\footnote{$A_{1}$ and $A_{2}$ are only allowed in the case of NO, since IO
places a lower bound on $(M_{\nu})_{ee}$ which controls the neutrinoless
double beta decay rate.} 

\begin{equation}
A_{1}:\,\begin{pmatrix}0 & 0 & X\\
0 & X & X\\
X & X & X
\end{pmatrix}\;,,\qquad A_{2}:\,\begin{pmatrix}0 & X & 0\\
X & X & X\\
0 & X & X
\end{pmatrix}\;,\label{eq:texturesA12}
\end{equation}
\begin{equation}
B_{1}:\,\begin{pmatrix}X & X & 0\\
X & 0 & X\\
0 & X & X
\end{pmatrix}\;,\qquad B_{2}:\,\begin{pmatrix}X & 0 & X\\
0 & X & X\\
X & X & 0
\end{pmatrix}\;,\label{eq:texturesB12}
\end{equation}
\begin{equation}
B_{3}:\,\begin{pmatrix}X & 0 & X\\
0 & 0 & X\\
X & X & X
\end{pmatrix}\;,\qquad B_{4}:\,\begin{pmatrix}X & X & 0\\
X & X & X\\
0 & X & 0
\end{pmatrix}\;,\label{eq:texturesB34}
\end{equation}
\begin{equation}
C:\,\begin{pmatrix}X & X & X\\
X & 0 & X\\
X & X & 0
\end{pmatrix}\;,\label{eq:textureC}
\end{equation}
and 8 textures which were already forbidden by data at the time they
were introduced:

\begin{equation}
D_{1}:\,\begin{pmatrix}X & X & X\\
X & 0 & 0\\
X & 0 & X
\end{pmatrix}\;,\qquad D_{2}:\,\begin{pmatrix}X & X & X\\
X & X & 0\\
X & 0 & 0
\end{pmatrix}\;,\label{eq:texturesD12}
\end{equation}
\begin{equation}
E_{1}:\,\begin{pmatrix}0 & X & X\\
X & 0 & X\\
X & X & X
\end{pmatrix}\;,\qquad E_{2}:\,\begin{pmatrix}0 & X & X\\
X & X & X\\
X & X & 0
\end{pmatrix}\;,\qquad E_{3}:\,\begin{pmatrix}0 & X & X\\
X & X & 0\\
X & 0 & X
\end{pmatrix}\;,\label{eq:texturesE13}
\end{equation}
\begin{equation}
F_{1}:\,\begin{pmatrix}X & 0 & 0\\
0 & X & X\\
0 & X & X
\end{pmatrix}\;,\qquad F_{2}:\begin{pmatrix}X & 0 & X\\
0 & X & 0\\
X & 0 & X
\end{pmatrix}\;,\qquad F_{3}:\,\begin{pmatrix}X & X & 0\\
X & X & 0\\
0 & 0 & X
\end{pmatrix}\;,\label{eq:texturesF13}
\end{equation}

Following a standard notation, in eqs.~(\ref{eq:texturesA12}\textendash \ref{eq:texturesF13})
we represent the position of the two zeroes of the complex symmetric
matrix in eq.~\eqref{eq:masses}. Two complex zeroes in the mass
matrix give 4 relations among the 9 real parameters entering $M_{\nu}$.
Depending on the texture, these relations will involve mainly the
well known oscillation parameters or the unknown non-oscillation parameters
$m_{\ell}$, $\alpha_{1}$ and $\alpha_{2}$ or a mixture of the two.
For instance $A_{1}$ texture gives all the 3 unknown parameters in
terms of the oscillation parameters, but in addition it gives a relation
between $\delta$ and the rest of the oscillation parameters which
can be tested against the experiment. On the other hand, $F_{1}$
gives several solutions, in one of them the masses are arbitrary but
$s_{12}=0$ and $s_{13}=0$ and therefore it cannot accommodate neutrino
oscillation data. Another solution, the only discussed usually in
the literature, gives $\alpha_{1}=\alpha_{2}=-2\delta$ and arbitrary
mixings, however, it requires exact degeneracy $m_{1}=m_{2}=m_{3}$,
therefore $\Delta_{21}=0$ and $\Delta_{31}=0$, and it is also excluded.\footnote{For a list of the analytical expressions for all textures, with conventions
slightly different from ours, see for instance \cite{Guo:2002ei}.}

It is important to remark that forbidden exact textures could become
allowed if the zeroes are only approximate, we will discuss some examples
in section \ref{sec:Approximate-texture-zeroes}.

\section{The method\label{sec:The-method}}

The NuFIT collaboration \cite{Esteban:2016qun,nufitweb} has fitted
all neutrino data on oscillations and made available the obtained
$\Delta\chi^{2}$ as a function of the six neutrino oscillations parameters
$s_{12}^{2}$, $s_{23}^{2}$, $s_{13}^{2}$, $\Delta_{21}$, $\Delta_{31}$
and $\delta$ (well, they offer marginalized $\Delta\chi^{2}$ for
each parameter individually and for all pairs of parameters). From
their results one can conclude that, in general, the correlations
are small except for the less known parameters $\delta$ and $s_{23}^{2}$
because the octant ambiguity. Therefore, from their data one can approximately
reconstruct the complete $\Delta\chi^{2}$ as 
\begin{equation}
\Delta\chi_{\nu f}^{2}\approx\Delta\chi^{2}(s_{23}^{2},\delta)+\Delta\chi^{2}(s_{13}^{2})+\Delta\chi^{2}(s_{12}^{2})+\Delta\chi^{2}(\Delta_{21})+\Delta\chi^{2}(\Delta_{31})-4\Delta\chi_{\mathrm{min}}^{2},\label{eq:chi2nufit}
\end{equation}
where the last term takes into account that the NuFIT collaboration
normalizes each of the different projections so that $\Delta\chi_{\mathrm{min}}^{2}=4.14$
in the case of IO, therefore we have to subtract four times $\Delta\chi_{\mathrm{min}}^{2}$
to keep the same normalization. We have checked that using this $\Delta\chi_{\nu f}^{2}$
one can reproduce reasonably well all correlation plots presented
in ref.~\cite{Esteban:2016qun,nufitweb}.

One way to see if the constraints imposed by the textures are compatible
with the data would be to vary all the oscillations parameters in
their allowed range (at 1$\sigma$, $2\sigma$, ...) and check if
the correlation is satisfied. Then, one can predict also the non-oscillation
parameters. This was done for instance in \cite{Singh:2016cbe,Singh:2016qcf}.
However, by using this method one does not take into account the correlations
of $\delta$ with $s_{23}^{2}$, which can be very important in some
cases.

Here we will use a different method, which has some advantages. We
will define a new $\chi^{2}$ that incorporates the constraints imposed
by the texture zeros with Lagrange multipliers
\begin{equation}
\chi^{2}=\Delta\chi_{\nu f}^{2}+\frac{1}{\lambda_{1}^{2}}|(M_{\nu})_{ab}|^{2}+\frac{1}{\lambda_{2}^{2}}|(M_{\nu})_{cd}|^{2}\,.\label{eq:chi2lambdas}
\end{equation}
For $\lambda_{1,2}\rightarrow+\infty$ one should recover the NuFIT
results while for $\lambda_{1,2}\rightarrow+0$ the constraints are
enforced maximally. The interpretation of $\lambda_{1}$ and $\lambda_{2}$
is also clear: the new terms only give an appreciable contribution
to the $\chi^{2}$ when $|(M_{\nu})_{ab}|>\lambda_{1}$ ($|(M_{\nu})_{cd}|>\lambda_{2}$).
Then, using this method we can also discuss approximate zeroes. Also,
in particular for a numerical treatment, one cannot set directly $\lambda_{1,2}$
to zero. For our purposes, as the rest of the parameters in the neutrino
mass matrix must be at least $1$ meV, it will be enough to take $\lambda_{1,2}\ll1$~meV.
To be definite, in our simulations we will take always $\lambda_{1}=\lambda_{2}=0.1$
meV and will check that the results do not change if we take smaller
values of the $\lambda$'s. Note that eq.~\eqref{eq:chi2lambdas}
has the standard $\chi^{2}$ interpretation of a measured Re$\{(M_{\nu})_{ab}\}$,Im$\{(M_{\nu})_{ab}\},=0\pm\lambda_{1}$
and similarly for $(M_{\nu})_{cd}$.

The method has another advantage because we can compute the $\chi^{2}$
at the minimum. This value will give us an indication of how well
the different textures are able to fit the data.

\section{Analysis of the allowed textures\label{sec:AllowedTextures}}

Following the method discussed above, we will analyze the different
allowed textures. Since $s_{12}^{2}$, $s_{13}^{2}$, $\Delta_{21}$,
$\Delta_{31}$ are well known from oscillation data, only textures
that can accommodate their values will be allowed. $\delta$ and $s_{23}^{2}$
are less known and there is more freedom to accommodate their values.
Thus it makes sense to represent the constraints imposed by the different
textures in the plane $s_{23}^{2}$\textendash $\delta$. For that
purpose we perform a simulation varying $s_{12}^{2}$, $s_{13}^{2}$,
$\Delta_{21}$, $\Delta_{31}$ in their allowed $3\sigma$ ranges
and find the region compatible with the different textures in the
$s_{23}^{2}$\textendash $\delta$ plane superposed to the NuFIT results
(this is shown, for instance on the left panel of figure~\ref{fig:A1}).
This gives a clear idea of the expected allowed regions when the texture
constraints are imposed to the NuFIT data according to eq.~\eqref{eq:chi2lambdas}.
It is important to remark that while the constraints imposed by all
the textures only depend on $\cos\delta$, and therefore are symmetric
with respect $\delta=180^{\circ}$, the global fit to neutrino oscillation
data is not, and this strongly constraints the overlap regions. 

The result of the complete fit is shown on the right panels ($s_{23}^{2}$\textendash $\delta$
allowed region, and the predictions for the non-oscillation parameters
$m_{\ell},\alpha_{1},\alpha_{2}$ against $\delta$). In all cases
contours correspond to two-dimensional 68.27\% 95.45\% 99.73\% C.L.
regions computed by minimization of the $\chi^{2}$ function in eq.~\eqref{eq:chi2lambdas}
for a fixed pair or parameters with respect to the rest of parameters
and then requiring\footnote{Notice that since we always subtract $\chi_{\mathrm{min}}$ to compute
C.L. regions, in the case of IO we do have a $1\sigma$ region, even
when the texture constraint is not imposed. This is different to what
is presented in the 2D plots by the NuFIT collaboration, where, in
the case of IO, the $\chi_{\mathrm{min}}=4.14$ relative to the NO
is not subtracted and therefore no $\text{1\ensuremath{\sigma}}$
region appears.} $\chi^{2}-\chi_{\mathrm{min}}^{2}<2.30,6.18,11.83$. For a more efficient
sampling of the parameter space we use a nested sampling algorithm
(MultiNest \cite{Feroz:2008xx,Feroz:2013hea}) and we do an explicit
$\chi^{2}$ minimization on the Markov Chain points. As discussed
above, we took $\lambda_{1}=\lambda_{2}=0.1$ meV and checked that,
in the case of the allowed textures, the results do not change if
we take $\lambda_{1}=\lambda_{2}=0.05$ meV. For $m_{\ell}$ we take
values in the range $0$\textendash $1000$ meV.

We have repeated this procedure for all the allowed textures. In section
\ref{sec:Approximate-texture-zeroes} we will discuss, in some cases,
how the constraints are relaxed if the textures are only approximate
by taking $\lambda_{1}=\lambda_{2}=5$ meV and also how forbidden
textures can become allowed if texture zeroes are only approximate,
by taking $\lambda_{1}=\lambda_{2}=1$~meV.

\subsection{A1 and A2 textures (only NO)}

\begin{figure}
\begin{centering}
\includegraphics[width=0.9\columnwidth]{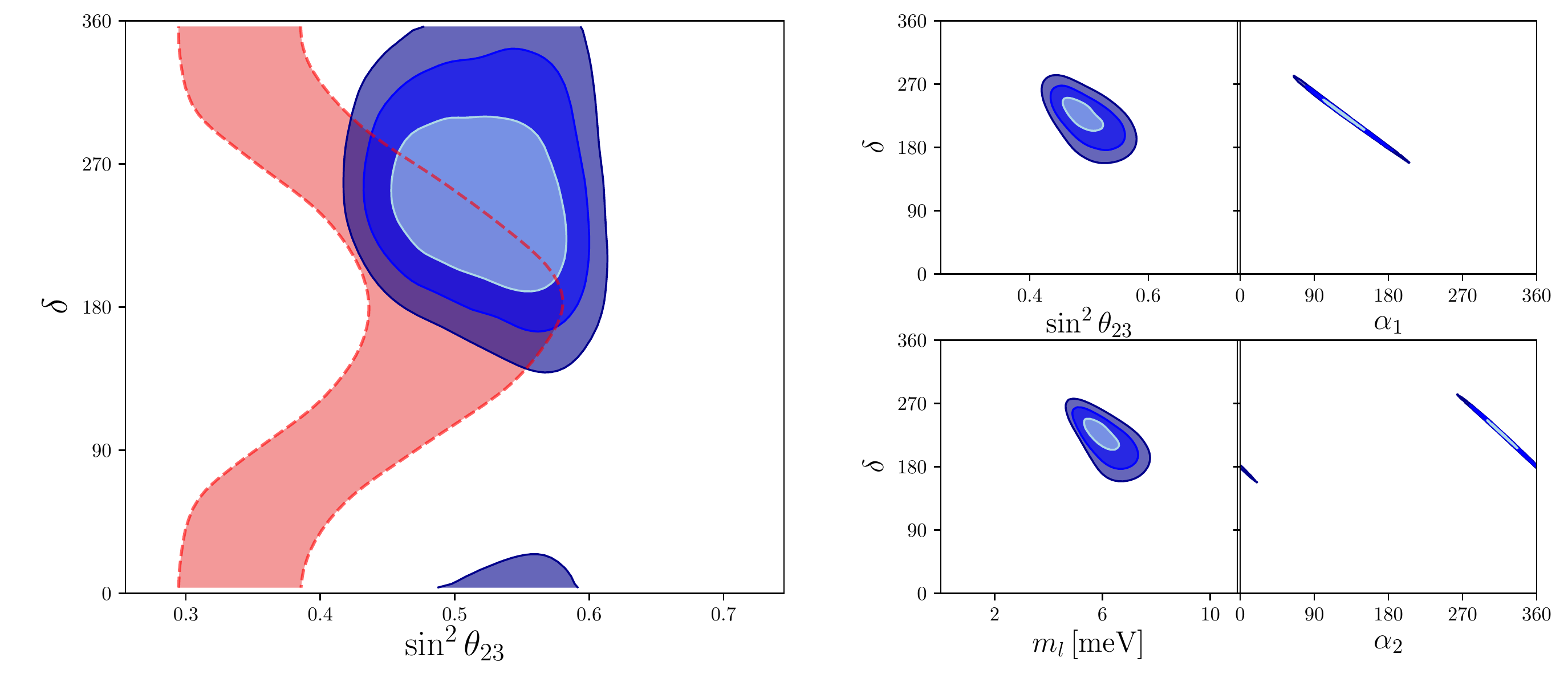}
\par\end{centering}

\begin{centering}
\includegraphics[width=0.9\columnwidth]{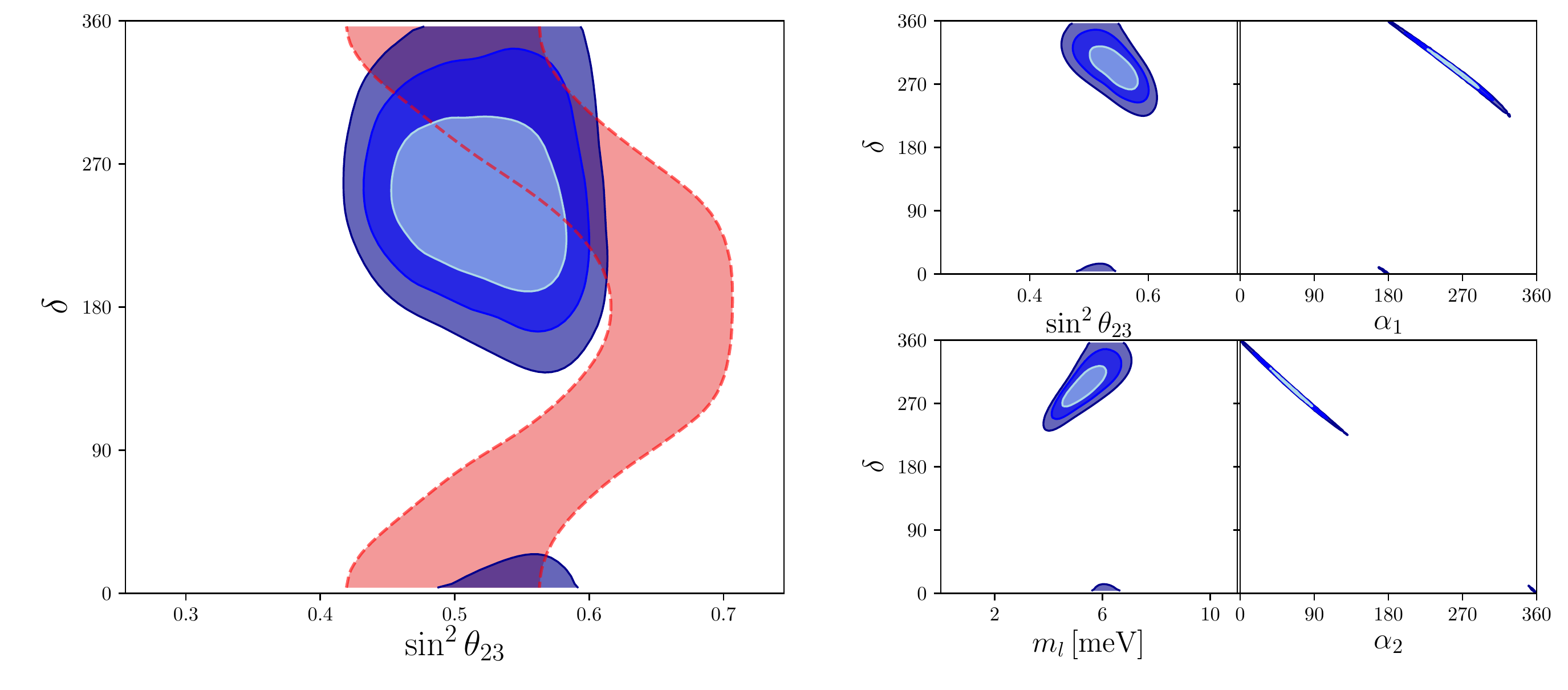}
\par\end{centering}

\caption{On the left of the panels we present the NuFIT results \cite{Esteban:2016qun,nufitweb},
in the $s_{23}^{2}$\textendash $\delta$ plane, for the global fit
to neutrino data (blue-gray coloured contours correspond to 68.27\%
95.45\% 99.73\% C.L. regions ) as compared with the prediction of
the textures, in red, obtained when the rest of the oscillation parameters
are varied in 3$\sigma$. On the right we present the new fit to the
data, as discussed in the text, when the constraints from the textures
are imposed. The upper panel is for the $A_{1}$ texture and the lower
one for the $A_{2}$ texture, which are only allowed in the NO case.\label{fig:A1}}
\end{figure}

Both $A_{1}$ and $A_{2}$ textures require $(M_{\nu})_{ee}=0$ which
is exactly the matrix element that controls the neutrinoless double
beta decay rate, $m_{\beta\beta}=(M_{\nu})_{ee}$. It is well known
the correlation between $m_{\beta\beta}$ and $m_{\ell}$ and the
fact that in the IO case $m_{\beta\beta}$ is bounded from below,
$m_{\beta\text{\ensuremath{\beta}}}\gtrsim10\,$meV \cite{Bilenky:1999wz,Vissani:1999tu},
therefore, $A_{1}$ and $A_{2}$ textures are only allowed in the
NO case.

The left panels of figure \ref{fig:A1} show clearly that in the case
of these two textures there are large regions of overlap between the
NuFIT results and the constraint imposed by the textures with $A_{1}$
giving some preference for slightly smaller values of $\delta$ while
$A_{2}$ prefers larger values.

On the right panels we present the constrained fit, eq.~\eqref{eq:chi2lambdas}.
The plane $s_{23}^{2}$\textendash $\delta$ obviously gives the overlap
regions shown on the left panels. In the rest of the plots we present
the predictions for the non-oscillations parameters, $m_{\ell}$,
$\alpha_{1}$ and $\alpha_{2}$ against $\delta$, which clearly show
the strong correlation between the Majorana phases $\alpha_{1},\alpha_{2}$
and $\delta$ and the fact that in these two textures the lightest
neutrino mass $m_{\ell}$ is predicted to be in a region around $5$~meV.

\subsection{B Textures}

\begin{figure}
\begin{centering}
\includegraphics[width=0.9\columnwidth]{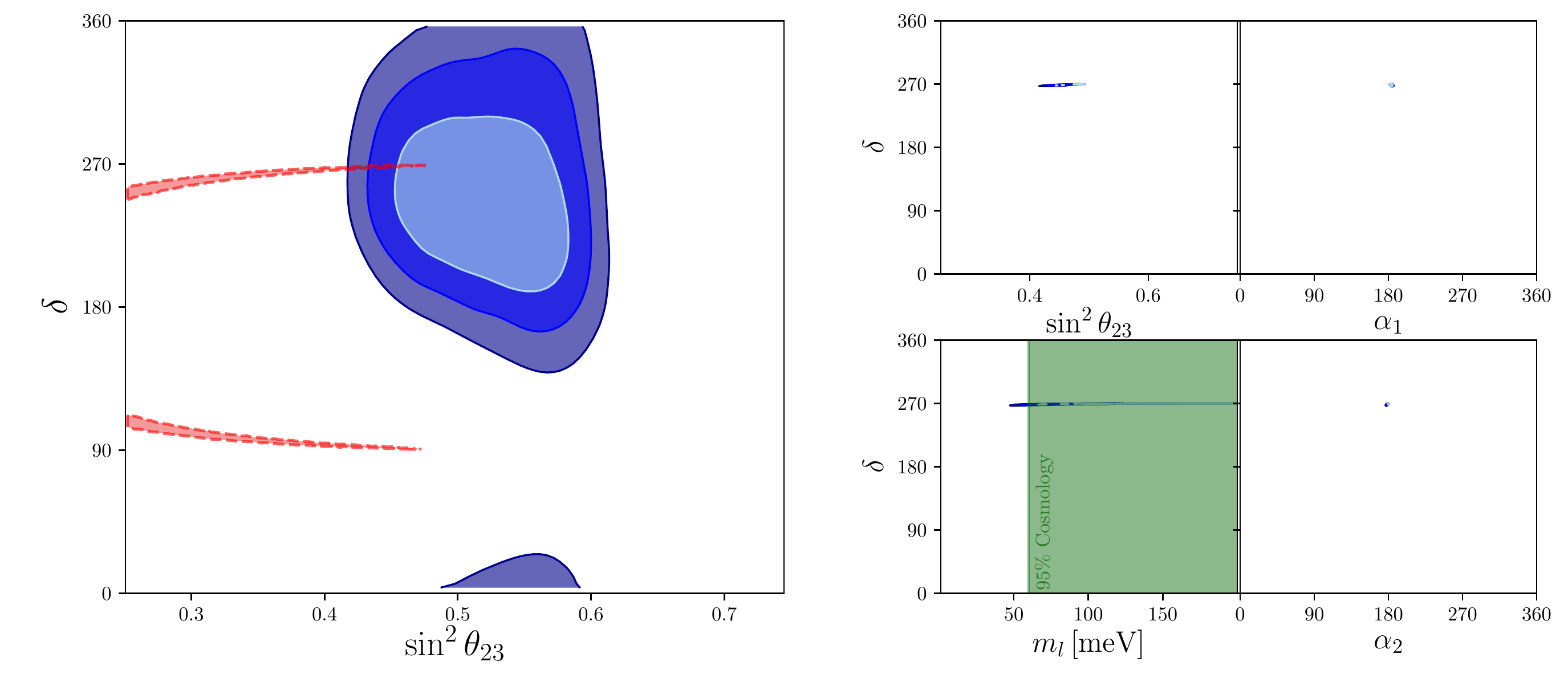}
\par\end{centering}

\begin{centering}
\includegraphics[width=0.9\columnwidth]{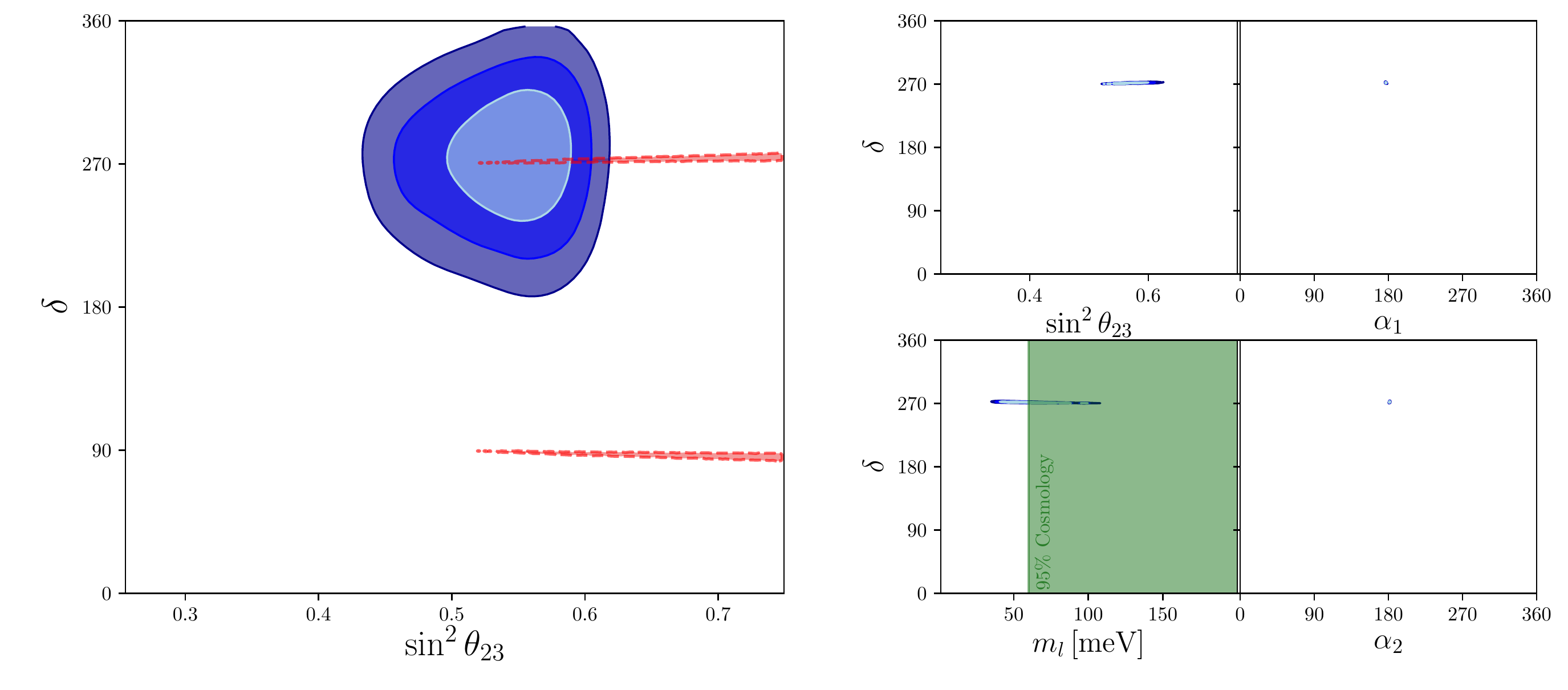}
\par\end{centering}

\caption{Same as figure~\ref{fig:A1} for the $B_{1}$ texture. Upper panels
for NO and lower ones for IO \label{fig:B1}. In the $\delta$\textendash $m_{\ell}$
plot we also present, for comparison, the bound on the lightest neutrino
mass obtained from Cosmology (we use $m_{\ell}<60$~meV \cite{Vagnozzi:2017ovm}). }
\end{figure}

Textures of type $B$ are all very similar and, taking into account
the ordering convention, we have eight of them. Basically they all
predict $\delta\sim270^{\circ}$ (or $\delta\sim90^{\circ}$ which
is strongly disfavoured by present fits), and $\alpha_{1}\sim\alpha_{2}\sim180^{\circ}$.
This is a consequence of the small value of $s_{13}^{2}$. They also
give a lower bound on the lightest neutrino mass, $m_{\ell}$, of
the order of $40$\textendash $50$~meV. Thus, we present in figure~\ref{fig:B1}
complete results only for texture $B_{1}$, in both NO and IO cases.

On the left panels of figure~\ref{fig:B1} we see the constraints
imposed by the $B_{1}$ texture for both the NO (above) and IO (below)
superimposed to the NuFIT contour plots in the plane $\text{\ensuremath{\delta}}$-$s_{23}^{2}$.
In the NO case $\delta$ is slightly below $270^{\circ}$ and $s_{23}^{2}<0.5$
while in the NO it is just the opposite ($\delta$ is slightly above
$270^{\circ}$ and $s_{23}^{2}>0.5$). Since central values of NuFIT
are slightly moved to higher values of $s_{23}^{2}$ in the case of
IO, in this case it seems there is a larger overlap region. Notice
that, as explained at the beginning of the section, to draw contours
we always use contours of $\chi-\chi_{\mathrm{min}}$ and, therefore,
these contours do not take into account that IO has a much larger
value of $\chi^{2}$ ($4.14$ relative to NO). 

On the right panels we present, as in figure~\ref{fig:A1}, the results
of the complete fit. The plots of $\delta$--$s_{23}^{2}$ just give
the tiny overlap region for $\delta\sim270^{\circ}$ and values of
$s_{23}^{2}<0.5$ in the NO case or $s_{23}^{2}>0.5$ in the IO one.
The $\alpha_{1},\alpha_{2}$ plots show they are basically fixed to
$\alpha_{1}\sim\alpha_{2}=180^{\circ}$. The plot of $\delta$ versus
$m_{\ell}$ is more interesting because it clearly shows that $m_{\ell}$
is bounded from below and can be rather large. For comparison we also
give, in green, the band forbidden by Cosmology (we take $m_{\text{\ensuremath{\ell}}}\leq(m_{1}+m_{2}+m_{3})/3\lesssim$
$60\,\mathrm{meV}$, ref.~\cite{Vagnozzi:2017ovm}, which includes
data from CMB and baryonic acoustic oscillations).

\begin{figure}
\begin{centering}
\includegraphics[width=1\columnwidth]{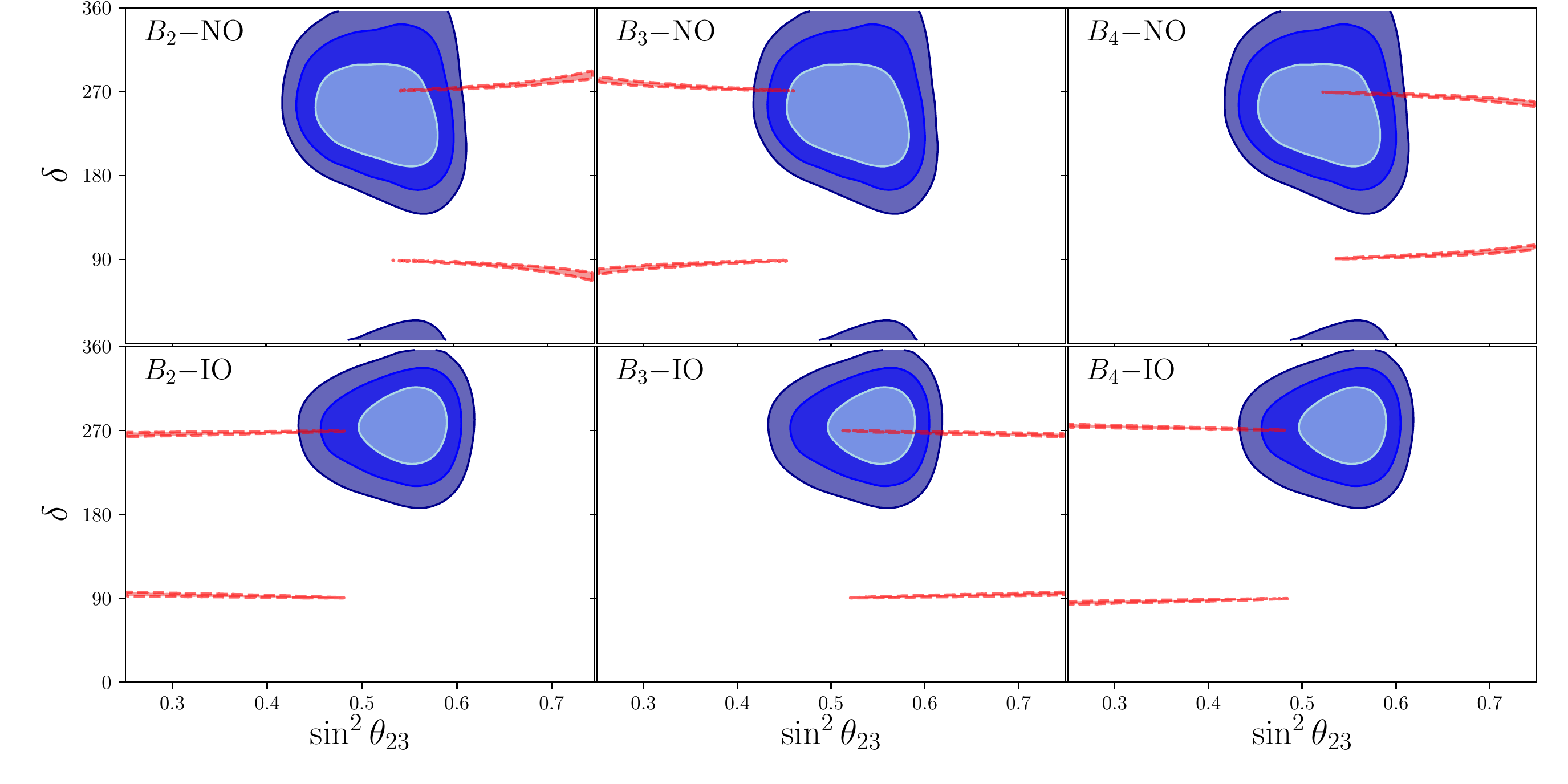}
\par\end{centering}

\caption{Same as the left panels of figure \ref{fig:B1} for the $B_{2}$,$B_{3}$
and $B_{4}$ textures. Above for in NO and below for IO.\label{fig:B2-B4}}
\end{figure}

For the rest of $B$ textures we present in figure~\ref{fig:B2-B4}
the region allowed by the textures on top of the NuFIT results in
the plane $s_{23}^{2}$\textendash $\delta$. We can see the small
differences between the different textures, $B_{1}$-NO,$B_{3}$-NO,$B_{2}$-IO,$B_{4}$-IO
require $s_{23}^{2}<0.5$ while $B_{1}$-IO,$B_{3}$-IO,$B_{2}$-NO,$B_{4}$-NO
require $s_{23}^{2}>0.5$. On the other hand $B_{1}$-NO,$B_{4}$-NO,$B_{2}$-IO,$B_{3}$-IO
give $\delta$ values a bit below $270^{\circ}$ while $B_{1}$-IO,$B_{4}$-IO,$B_{2}$-NO,$B_{3}$-NO
give $\delta$ a bit above $270^{\circ}$. The exact bands allowed
at the $3\sigma$ level, together with the limits on the masses, are
presented in table~\ref{tab:chi2}.

\subsection{C Texture}

\begin{figure}
\begin{centering}
\includegraphics[width=0.9\columnwidth]{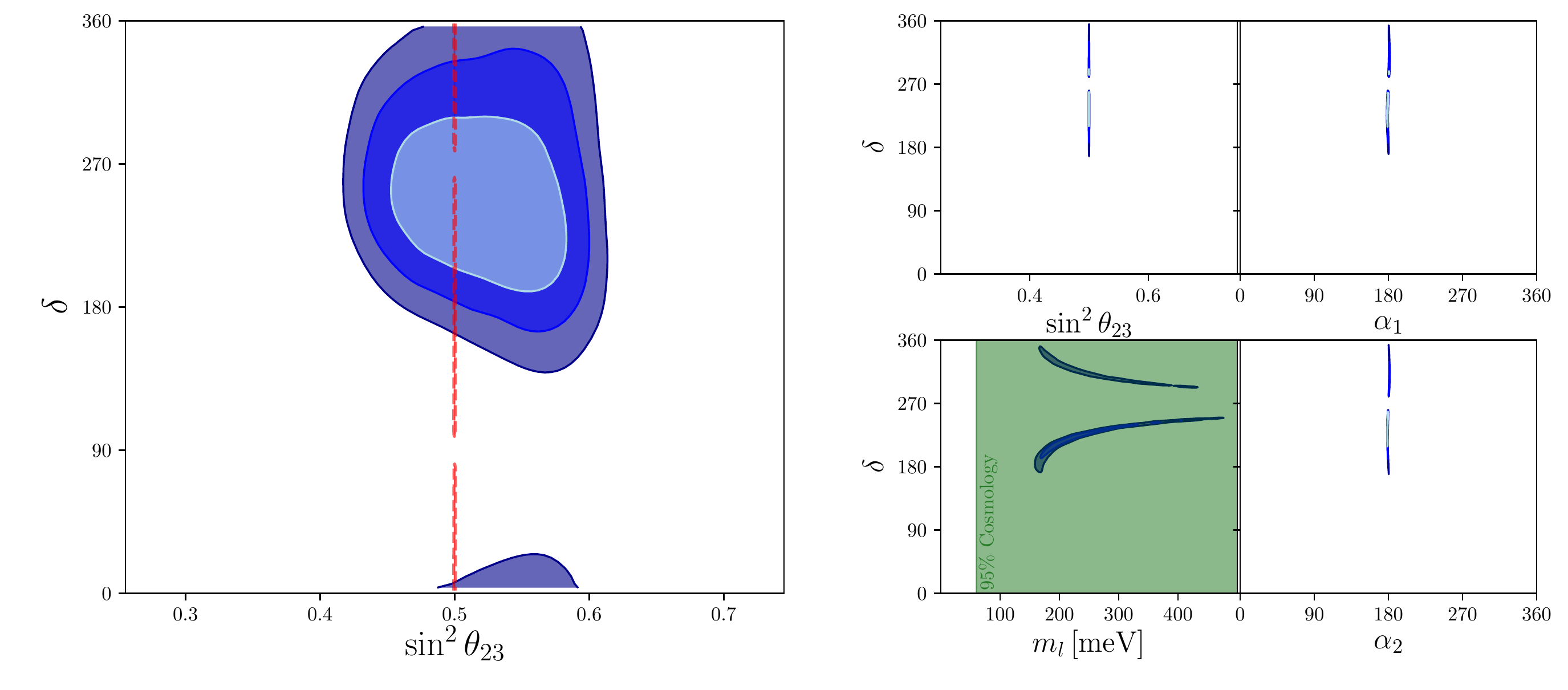}
\par\end{centering}

\begin{centering}
\includegraphics[width=0.9\columnwidth]{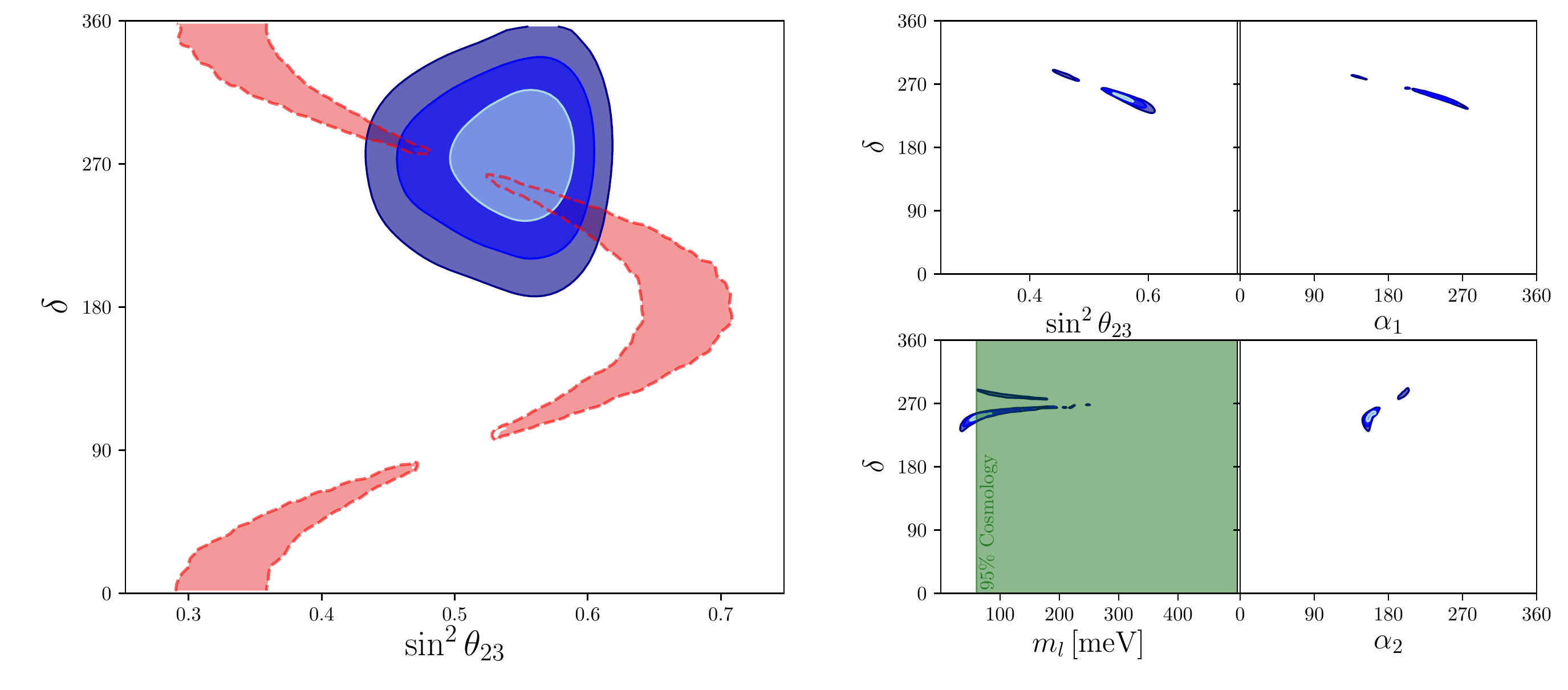}
\par\end{centering}

\caption{Same as figure~\ref{fig:B1} for the $C$ texture. In the upper panels
for NO and the lower ones for IO.\label{fig:C}}
\end{figure}

Texture $C$, represented in figure \ref{fig:C}, is probably the
most peculiar of the textures since it divides the space of parameters
in two disjoint regions according to the exact value of $\delta$
(this is clearly seen in the $\delta$\textendash $m_{\ell}$ plot)
. 

In the case of NO it predicts $s_{23}^{2}\simeq0.5$ with a high degree
of precision (see for instance \cite{Grimus:2004az}) and forbids
a small region around $\delta\sim270^{\circ}$ (well, it requires
very large values of $m_{\ell}$ to reach it). Since the last NuFIT
results seem to favour values around $s_{23}^{2}=0.5$, as we will
see in table \ref{tab:chi2}, this texture gives one of the lowest
values of the $\chi^{2}$, but this is at the cost of very large values
of $m_{\ell}$, which, as shown in the $\delta$\textendash $m_{\ell}$
plot, can be in conflict with Cosmology data. For the Majorana phases
it gives $\alpha_{1}\sim\alpha_{2}\sim180^{\circ}$ (see table \ref{tab:chi2}
for the exact values). 

The IO case is even more peculiar. It forbids small regions around
$\delta=270^{\circ}$ and $s_{23}^{2}$ around $0.5$, and this is
also translated into the possible values of $\alpha_{1}$ and $\alpha_{2}$.
On the other hand, even though it also gives a lower bound on $m_{\ell}$,
there is still some space to make it compatible with Cosmology.

\subsection{Best fit parameters}

In table \ref{tab:chi2} we give the $3\sigma$ bands for the relevant
parameters ($\Delta_{21}$, $\Delta_{31}$, $s_{12}$, $s_{13}$ are
within the standard oscillation fit ranges) in the different allowed
textures in both the NO and IO cases. We also present allowed bands
for the lightest neutrino mass $m_{\text{\ensuremath{\ell}}}$, the
Cosmology mass $m_{\mathrm{cos}}=m_{1}+m_{2}+m_{3}$ and the effective
mass relevant for neutrinoless double beta decay $m_{\beta\beta}=|(M_{\nu})_{ee}|$.
We also give the $\chi^{2}$ on the best fit parameters. Finally,
to see the impact of the Cosmology bound, $m_{\ell}<60\,$meV, in
the last column we also present the $\chi^{2}$ values obtained when
$m_{\ell}<60\,$meV is imposed. Following the NuFIT collaboration,
in the in the IO case we have included the value of the minimum, $4.14$,
relative to the absolute minimum of NuFIT which happens for NO. However
to compute the 3$\sigma$ bands we take, as usual, $\chi-\chi_{\mathrm{min}}=9$.

All textures considered, give $\chi^{2}$ values around 1 ($B_{2,4}$\textendash IO
which give slightly larger values). This is really interesting since,
as discussed in the introduction, two-zero neutrino textures depend
on only five real parameter from the six oscillation parameters.

On the other hand, looking at figs.~\ref{fig:B1}--\ref{fig:C} one
can see that in the $s_{23}^{2}-\delta$ plane the constraints for
$B's$ and $C$\textendash NO are basically lines, so the amount of
parameter space is very small. This can be measured using a Bayesian
estimator, like the Bayes factor, but to compute it is technically
complicated and has also its own conceptual problems because the comparison
depends strongly on the volume of the priors and their parametrization,
thus in this paper we decided to present only the $\chi^{2}$ values
at the minimum. 

\begin{table}
\begin{centering}
\begin{tabular}{|c|c|c|c|c|c|c|c|c|c|}
\hline 
 & $m_{\ell}$ & $m_{\mathrm{cos}}$ & $m_{\beta\beta}$ & $s_{23}^{2}$ & $\delta$ & $\alpha_{1}$ & $\alpha_{2}$ & $\chi^{2}$ & $\chi_{m_{\ell}<60}^{2}$\tabularnewline
\hline 
\hline 
$A_{1}$\textendash NO & 4.2\textendash 7.8 & 64-70 & 0\textendash 0.3 & 0.42\textendash 0.59  & 154\textendash 290  & 56\textendash 210 & 256\textendash 383  & 0.8 & 0.8\tabularnewline
\hline 
$A_{2}$\textendash NO & 3.4\textendash 7.1 & 62-69 & 0\textendash 0.3 & 0.45\textendash 0.62 & 217\textendash 369 & 169\textendash 334  & -8\textendash 139  & 1.9 & 1.9\tabularnewline
\hline 
$B_{1}$\textendash NO & >47  & >170 & 50\textendash 245 & 0.42\textendash 0.50  & 267\textendash 270  & 180\textendash 187  & 177\textendash 180 & 0.7 & 4\tabularnewline
\hline 
$B_{1}$\textendash IO & >37 & >165 & 62\textendash 195 & 0.50\textendash 0.62 & 269\textendash 271 & 175-180 & 180\textendash 182 & 4.2 & 4.2\tabularnewline
\hline 
$B_{2}$\textendash NO & >39  & >147 & 41\textendash 202 & 0.50\textendash 0.61 & 270\textendash 274  & 170\textendash 180 & 180\textendash 184 & 0.7 & 1.5\tabularnewline
\hline 
$B_{2}$\textendash IO & >48 & >205 & 74\textendash 315 & 0.43\textendash 0.50 & 269\textendash 271 & 180\textendash 183 & 179\textendash 180 & 6.2 & 12\tabularnewline
\hline 
$B_{3}$\textendash NO & >50 & >179 & 53\textendash 249 & 0.42\textendash 0.50  & 270\textendash 273  & 176\textendash 180  & 180\textendash 182  & 0.7 & 5\tabularnewline
\hline 
$B_{3}$\textendash IO & >40 & >172 & 64\textendash 266 & 0.50\textendash 0.62 & 268\textendash 271 & 180\textendash 186 & 177-180 & 4.2 & 4.4\tabularnewline
\hline 
$B_{4}$ \textendash NO & >41 & >153 & 43\textendash 206 & 0.50\textendash 0.61 & 266\textendash 270 & 180\textendash 186 & 178\textendash 180 & 0.7 & 1.7\tabularnewline
\hline 
$B_{4}$\textendash IO & >56  & >212 & 76\textendash 334 & 0.43\textendash 0.50  & 269\textendash 271  & 176\textendash 180  & 180\textendash 182  & 6.2 & 13\tabularnewline
\hline 
\multirow{2}{*}{$C$\textendash NO} & >159 & >484 & >151 & \multirow{2}{*}{0.50 } & 175\textendash 262  & 178\textendash 180  & 178-180  & 0.2  & >1000\tabularnewline
 & >167 &  &  &  & 278\textendash 346 & 180\textendash 182 & 180-182 & 1.1 & >1000\tabularnewline
\hline 
\multirow{2}{*}{$C$\textendash IO} & >35 & >155 & >34 & 0.51\textendash 0.61 & 231\textendash 269  & 186\textendash 281 & 151\textendash 178  & 4.8 & 5.1\tabularnewline
 & >67 &  &  & 0.44\textendash 0.49 & 273\textendash 289 & 120\textendash 168  & 185\textendash 202 & 6.7 & 13\tabularnewline
\hline 
\end{tabular}
\par\end{centering}

\caption{99.73\% C.L. results for the fits of the different textures. All masses
are given in meV and all angles in degrees taken in the range $[0,360^{\circ}]$,
except in some cases in which to avoid disjoint bands we have enlarged
the region slightly below $0^{\circ}$ or above $360^{\circ}$. IO
$\chi^{2}$ already include the $4.1$4 relative to NO obtained by
the NuFIT collaboration but it is subtracted to compute the bands.
We present also the $\chi^{2}$ values obtained when the Cosmology
bound $m_{\ell}<60$~meV is imposed.\label{tab:chi2} }
\end{table}

\section{Approximate texture zeroes\label{sec:Approximate-texture-zeroes}}

Above we have considered only allowed textures in the limit in which
the texture zeroes are exact. In specific models the texture zeros
come from symmetries which are slightly broken or are consequences
of the dynamics (zeroes could arise only at some order of perturbation
theory or be proportional to some small couplings). Moreover, one
expects, that in some cases, radiative corrections will fill the texture
zeroes with small quantities~\cite{Hagedorn:2004ba}. Then, it makes
sense to ask how stable are the conclusions of our analysis against
small perturbations. On the other hand, it is possible that textures
that were considered excluded, if they are exact, become allowed if
the zeroes are only approximate. The method proposed in this paper
makes it trivial to discuss these problems. 

\begin{figure}
\begin{centering}
\includegraphics[width=0.6\columnwidth]{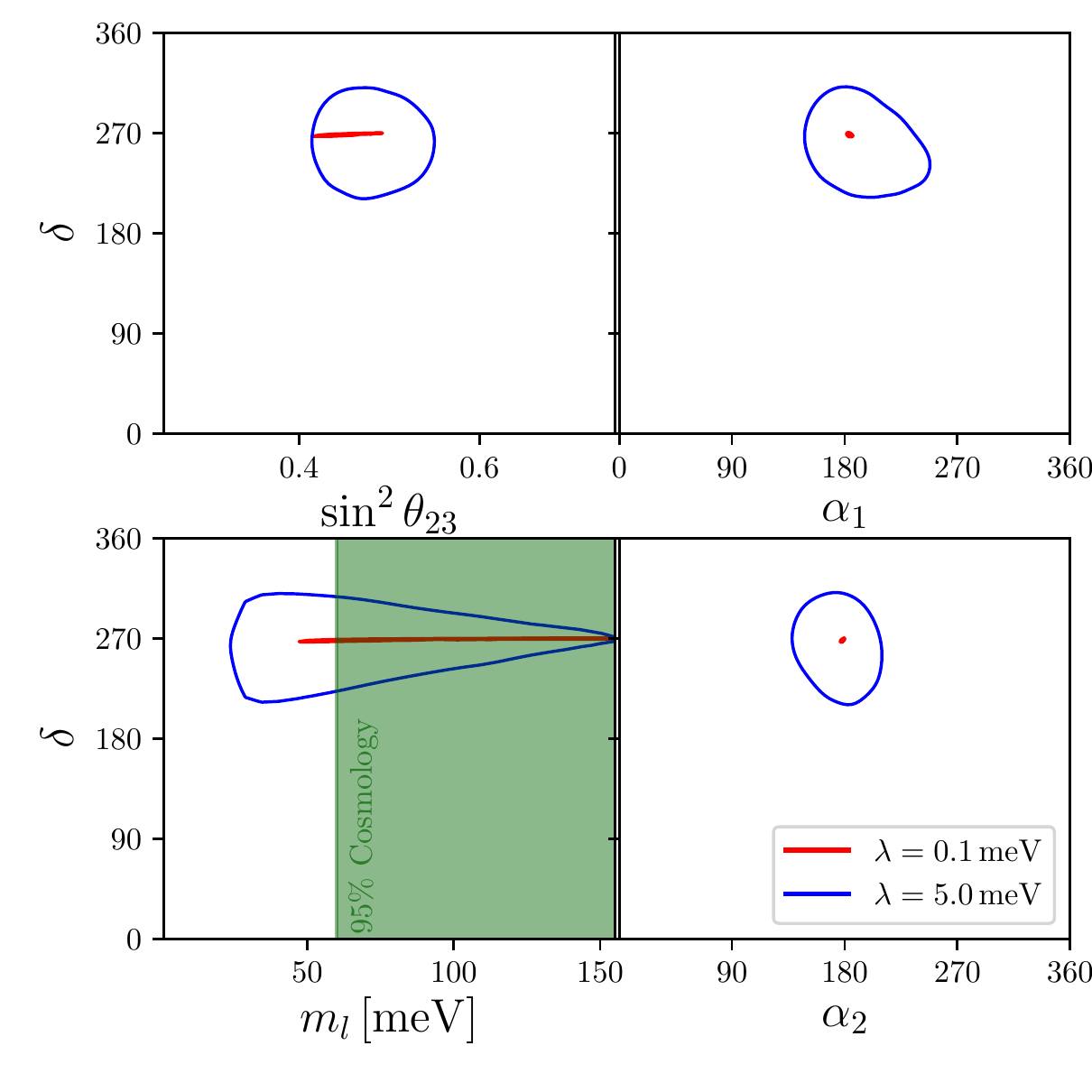}
\par\end{centering}

\caption{Allowed $3\sigma$ regions in the case of the $B_{1}$\textendash NO
texture when the texture zeroes are only approximate.\label{fig:approximateB1}}
\end{figure}

To answer the first question we have considered the texture $B_{1}$\textendash NO
and studied how the parameter space changes when we move from $\lambda=0.1$~meV
to $\lambda=5$~meV. This is a typical example with a very constrained
parameter space ($\delta$, $\alpha_{1}$ and $\alpha_{2}$ are basically
fixed). In figure~\ref{fig:approximateB1} we compare the allowed
parameter space in these two cases and see that when the texture zeroes
are not exact the available parameter space increase enormously but
still the main predictions of the texture remain. 

\begin{figure}
\begin{centering}
\includegraphics[width=0.6\columnwidth]{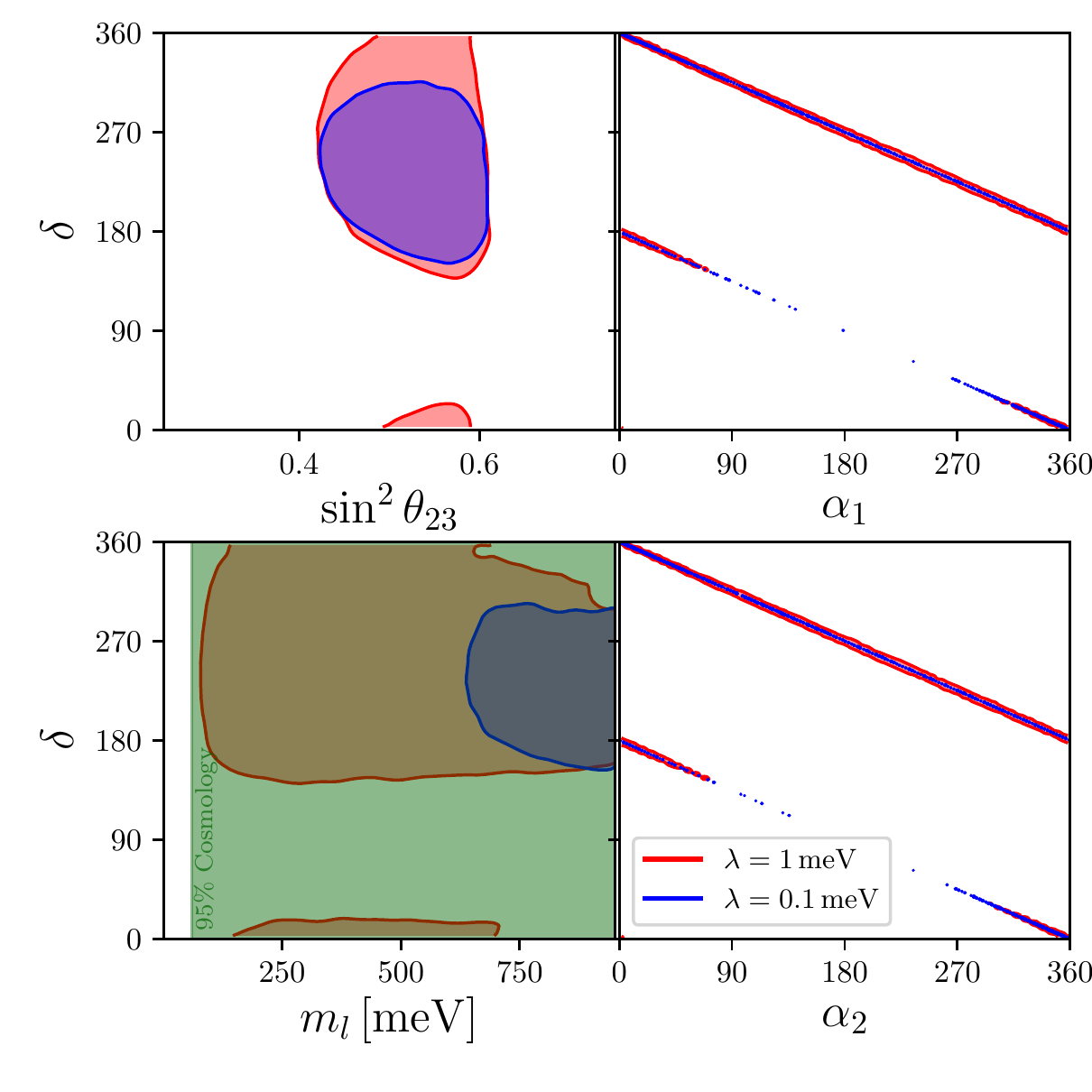}
\par\end{centering}

\caption{Allowed $3\sigma$ regions in the case of the ``excluded'' $F_{1}$\textendash NO
texture when the texture zeroes are only approximate.\label{fig:approximateF1-NO}}
\end{figure}

By using our method we could easily check quantitatively at which
level the forbidden textures are excluded and how stable is this exclusion
when the zeroes are not exact. Thus, we first minimized the $\chi^{2}$
in eq.~\eqref{eq:chi2lambdas} for $\lambda=0.1$~meV. We found
that all textures, except $F$-type textures, give very large $\chi^{2}$
(larger than 50 for all of them and in some cases, $D$\textendash NO,
over $1000$). However, $F$-textures, give somehow lower values,
and in particular $F_{1}$\textendash $NO$ gives a fit with $\chi^{2}$
below $10$. These results are even more clear when we increase $\lambda$
from $0.1$~meV to $1$~meV, in which case $\chi^{2}$ as low as
$0.5$ can be obtained ($F_{1}$\textendash NO).\footnote{Notice that precisely $F$-type textures are expected to receive larger
radiative corrections~\cite{Hagedorn:2004ba}.} while the rest of the forbidden textures still give a large $\chi^{2}$.
To understand this result it is useful to see why exact $F$-textures
are forbidden. Take the case of $F_{1}$ for instance, its mass matrix
has two zeroes at the elements $(M_{\nu})_{12}$ and $(M_{\nu})_{13}$
and it is block-diagonal with $(M_{\nu})_{23}$ as the only non-trivial
non-diagonal element. Then, one would conclude that $s_{12}=s_{13}=0$,
which, of course, cannot accommodate neutrino oscillation data. \footnote{This trivial and natural solution has been dismissed in works that
use the method of ref.~\cite{Xing:2002ta,Guo:2002ei}.} However, this is not the only solution since, by taking the parametrization
in eqs.~(\ref{eq:masses}\textendash \ref{eq:UPMNS}), one can easily
see that if $m_{1}=m_{2}=m_{3}$ there are also solutions with arbitrary
mixings (see for instance \cite{Guo:2002ei}). This solution also
implies that $\alpha_{1}=\alpha_{2}=-2\delta$ (or $\alpha_{1}=\alpha_{2}=2k\pi-2\delta$~$k=\pm1,\pm2,\cdots$).
However, if masses are degenerate, $\Delta_{21}=\Delta_{31}=0$, and
oscillation data cannot be accommodated either. In the method we are
proposing all possible solutions are included automatically. Thus,
in figure~\ref{fig:approximateF1-NO} we present results, in the
case of $F_{1}$\textendash NO, for $\lambda=0.1$~meV and $\lambda=1$~meV.
All the oscillation parameters, including $\Delta_{31}$ and $\Delta_{21}$,
can be adjusted easily in the two cases although for $\lambda=0.1$~meV
$\chi^{2}$ is somehow larger but still below $9$. Moreover, to fit
the data, large values of $m_{\ell}$ are required (see figure~\ref{fig:approximateF1-NO}).
If we take $\lambda=1$~meV the fit is much improved ($\chi^{2}$
below 1) and allows for much smaller values of $m_{\ell}$. 

One interesting point is that the correlation between phases $\alpha_{1}=\alpha_{2}=-2\delta$,
remains in spite of the non-exact texture zeros. This result can be
understood by using standard degenerate perturbation theory: if the
exact texture produces a degenerate spectrum and we introduce a small
perturbation, it will shift the eigenvalues by a small quantity but
the mixings, given by the eigenvectors which diagonalize the perturbation,
will not be suppressed and can be as large as needed to fit the data.
In the case of $F_{1}$\textendash NO, for $M_{12,13}\ll m_{3}$ one
typically finds

\[
\Delta_{31}\propto\frac{1}{s_{13}}m_{3}|M_{12,13}|\;,\quad\Delta_{21}\propto m_{3}|M_{12,13}|
\]
with coefficients which depend on $s_{12}$ and $s_{23}$. This shows
a natural enhancement of $\Delta_{31}$ with respect to $\Delta_{21}$
due to the smallness of $s_{13}$. Moreover, since $\Delta_{31,21}$
are fixed by oscillations, and we are requiring $|M_{12,13}|<\lambda$,
it is clear that, in general, smaller $\lambda$'s will require larger
$m_{3}$ to fit the data, as clearly seen in figure~\ref{fig:approximateF1-NO}.
On the other hand, for the phases we have 
\[
e^{i(\alpha_{1}+2\delta)}\approx e^{i(\alpha_{2}+2\delta)}\approx1+\mathcal{O}\left(\frac{M_{12,13}}{m_{3}}\right)
\]
which explains the strong correlation of the Majorana phases with
$\delta$ even when the texture is only approximate.

\section{Conclusions\label{sec:Conclusions}}

We have introduced a new method, based on a $\chi^{2}$ analysis with
constraints, to analyze numerically possible relations among the elements
of the fermion mass matrices. As an example, we have applied it to
the case of two-zero neutrino textures. The method has allowed us
to disentangle the available parameter space, give correlation plots
among parameters and confidence level bands without any algebraic
work. We have also compared the different textures according the minimum
$\chi^{2}$ they can give. 

In the case of the known ``allowed'' textures, $A,B,C$ we have
seen that, although in some cases $B$ and $C$ textures offer a fit
with smaller values of $\chi^{2}$, $A$ textures are favoured with
respect the rest of the allowed textures because:

-They have a larger parameter space: the highly restricted values
of $\delta$ in the other textures, especially $B$ textures, will
make it difficult to accommodate them if the oscillation data becomes
more precise. In fact, already now textures $B_{2}$--IO and $B_{4}$--IO
have no overlapping region at the $\text{1\ensuremath{\sigma}}$ level
with the last NuFIT results, which is manifested in a $\chi^{2}$
above $6$ (this already takes into account the IO $\chi^{2}$ minimum
value of 4.14 relative to NO).

-All textures, except $A$\textendash type textures, require large
values of the lightest neutrino mass, $m_{\ell}\gtrsim40$~meV, in
particular $m_{\ell}\gtrsim160$~meV in the $C$-NO texture. This
can be in tension with Cosmology, which at present requires $m_{\ell}<60\,$meV.
But, of course, there could be some, still unknown mechanism, that
could make Cosmology data compatible with larger neutrino masses. 

On the other hand, neutrinoless double beta decay experiments will
provide another test of the textures. If it is found in the next round
of experiments $A_{1}$ and $A_{2}$ textures, at least if they are
exact, will be excluded since they require $m_{\beta\beta}=0$. 

We have also discussed approximate texture zeros. We found that in
the case of the allowed textures the general conclusions are not changed
if the zero-matrix elements are below $1$~meV, although in the case
of textures $B$, which have a strongly constrained parameter space,
it is enlarged if the zeros are just $5$~meV. More importantly,
we have also analyzed the case of forbidden textures by taking matrix
elements below $1$~meV. In general, all forbidden textures remain
forbidden (they give values of $\chi^{2}$ above $50$). However,
textures of type $F$, could become allowed with $\chi^{2}$ which
are below $1$, in particular, in the case of $F_{1}$\textendash NO
and $F_{3}$-NO. 

Finally we have shown that the numeric method proposed in this paper
is a good complement of the analytic studies to study the relations
between Yukawa couplings/mass matrices imposed by symmetries or the
flavour structure of the theory. The method incorporates naturally
correlations among measured parameters, allows us to compute the available
parameter space and provides a standard $\chi^{2}$ comparative test
of how well the different models can accommodate the experimental
data. It also generalizes trivially to the case in which the relations
among parameters are only approximate.
\begin{acknowledgments}
This work is partially supported by the Spanish MINECO under grants
FPA2014-54459-P, FPA2014-57816-P, FPA2016-76005-C2-1-P, FPA2017-84543-P,
2017-SGR-929, by the Severo Ochoa Excellence Program under grant SEV-2014-0398
and by the \textquotedblleft Generalitat Valenciana\textquotedblright{}
under grants GVPROMETEOII2014-087, GVPROMETEOII/2014/050. J.S. is
also supported by the EU Networks FP10 ITN ELUSIVES (H2020-MSCA-ITN-2015-674896)
and INVISIBLESPLUS (H2020-MSCA-RISE-2015-690575). 
\end{acknowledgments}

\bibliographystyle{JHEP}
\bibliography{references}

\providecommand{\href}[2]{#2}\begingroup\raggedright\begin{thebibliography}{10}

\bibitem{Wilczek:1977uh}
F.~Wilczek and A.~Zee, \emph{{Discrete Flavor Symmetries and a Formula for the
  Cabibbo Angle}},
  \href{http://dx.doi.org/10.1016/0370-2693(77)90403-8}{\emph{Phys. Lett.} {\bf
  B70} (1977) 418}.

\bibitem{Fritzsch:1977za}
H.~Fritzsch, \emph{{Calculating the Cabibbo Angle}},
  \href{http://dx.doi.org/10.1016/0370-2693(77)90408-7}{\emph{Phys. Lett.} {\bf
  B70} (1977) 436--440}.

\bibitem{Ramond:1993kv}
P.~Ramond, R.~G. Roberts and G.~G. Ross, \emph{{Stitching the Yukawa quilt}},
  \href{http://dx.doi.org/10.1016/0550-3213(93)90159-M}{\emph{Nucl. Phys.} {\bf
  B406} (1993) 19--42}, [\href{http://arxiv.org/abs/hep-ph/9303320}{{\tt
  hep-ph/9303320}}].

\bibitem{Frampton:2002yf}
P.~H. Frampton, S.~L. Glashow and D.~Marfatia, \emph{{Zeroes of the neutrino
  mass matrix}},
  \href{http://dx.doi.org/10.1016/S0370-2693(02)01817-8}{\emph{Phys. Lett.}
  {\bf B536} (2002) 79--82}, [\href{http://arxiv.org/abs/hep-ph/0201008}{{\tt
  hep-ph/0201008}}].

\bibitem{Grimus:2004hf}
W.~Grimus, A.~S. Joshipura, L.~Lavoura and M.~Tanimoto, \emph{{Symmetry
  realization of texture zeros}},
  \href{http://dx.doi.org/10.1140/epjc/s2004-01896-y}{\emph{Eur. Phys. J.} {\bf
  C36} (2004) 227--232}, [\href{http://arxiv.org/abs/hep-ph/0405016}{{\tt
  hep-ph/0405016}}].

\bibitem{Dev:2011jc}
S.~Dev, S.~Gupta and R.~R. Gautam, \emph{{Zero Textures of the Neutrino Mass
  Matrix from Cyclic Family Symmetry}},
  \href{http://dx.doi.org/10.1016/j.physletb.2011.06.046}{\emph{Phys. Lett.}
  {\bf B701} (2011) 605--608}, [\href{http://arxiv.org/abs/1106.3451}{{\tt
  1106.3451}}].

\bibitem{Zee:1985id}
A.~Zee, \emph{{Quantum Numbers of Majorana Neutrino Masses}},
  \href{http://dx.doi.org/10.1016/0550-3213(86)90475-X}{\emph{Nucl. Phys.} {\bf
  B264} (1986) 99--110}.

\bibitem{Babu:1988ki}
K.~S. Babu, \emph{{Model of 'Calculable' Majorana Neutrino Masses}},
  \href{http://dx.doi.org/10.1016/0370-2693(88)91584-5}{\emph{Phys. Lett.} {\bf
  B203} (1988) 132--136}.

\bibitem{delAguila:2011gr}
F.~del Aguila, A.~Aparici, S.~Bhattacharya, A.~Santamaria and J.~Wudka,
  \emph{{A realistic model of neutrino masses with a large neutrinoless double
  beta decay rate}},
  \href{http://dx.doi.org/10.1007/JHEP05(2012)133}{\emph{JHEP} {\bf 05} (2012)
  133}, [\href{http://arxiv.org/abs/1111.6960}{{\tt 1111.6960}}].

\bibitem{delAguila:2012nu}
F.~del Aguila, A.~Aparici, S.~Bhattacharya, A.~Santamaria and J.~Wudka,
  \emph{{Effective Lagrangian approach to neutrinoless double beta decay and
  neutrino masses}},
  \href{http://dx.doi.org/10.1007/JHEP06(2012)146}{\emph{JHEP} {\bf 06} (2012)
  146}, [\href{http://arxiv.org/abs/1204.5986}{{\tt 1204.5986}}].

\bibitem{Alcaide:2017xoe}
J.~Alcaide, D.~Das and A.~Santamaria, \emph{{A model of neutrino mass and dark
  matter with large neutrinoless double beta decay}},
  \href{http://dx.doi.org/10.1007/JHEP04(2017)049}{\emph{JHEP} {\bf 04} (2017)
  049}, [\href{http://arxiv.org/abs/1701.01402}{{\tt 1701.01402}}].

\bibitem{Branco:2007nn}
G.~C. Branco, D.~Emmanuel-Costa, R.~Gonzalez~Felipe and H.~Serodio, \emph{{Weak
  Basis Transformations and Texture Zeros in the Leptonic Sector}},
  \href{http://dx.doi.org/10.1016/j.physletb.2008.10.059}{\emph{Phys. Lett.}
  {\bf B670} (2009) 340--349}, [\href{http://arxiv.org/abs/0711.1613}{{\tt
  0711.1613}}].

\bibitem{Guo:2002ei}
W.-l. Guo and Z.-z. Xing, \emph{{Implications of the KamLAND measurement on the
  lepton flavor mixing matrix and the neutrino mass matrix}},
  \href{http://dx.doi.org/10.1103/PhysRevD.67.053002}{\emph{Phys. Rev.} {\bf
  D67} (2003) 053002}, [\href{http://arxiv.org/abs/hep-ph/0212142}{{\tt
  hep-ph/0212142}}].

\bibitem{Dev:2006qe}
S.~Dev, S.~Kumar, S.~Verma and S.~Gupta, \emph{{Phenomenology of two-texture
  zero neutrino mass matrices}},
  \href{http://dx.doi.org/10.1103/PhysRevD.76.013002}{\emph{Phys. Rev.} {\bf
  D76} (2007) 013002}, [\href{http://arxiv.org/abs/hep-ph/0612102}{{\tt
  hep-ph/0612102}}].

\bibitem{Fritzsch:2011qv}
H.~Fritzsch, Z.-z. Xing and S.~Zhou, \emph{{Two-zero Textures of the Majorana
  Neutrino Mass Matrix and Current Experimental Tests}},
  \href{http://dx.doi.org/10.1007/JHEP09(2011)083}{\emph{JHEP} {\bf 09} (2011)
  083}, [\href{http://arxiv.org/abs/1108.4534}{{\tt 1108.4534}}].

\bibitem{Meloni:2012sx}
D.~Meloni and G.~Blankenburg, \emph{{Fine-tuning and naturalness issues in the
  two-zero neutrino mass textures}},
  \href{http://dx.doi.org/10.1016/j.nuclphysb.2012.10.011}{\emph{Nucl. Phys.}
  {\bf B867} (2013) 749--762}, [\href{http://arxiv.org/abs/1204.2706}{{\tt
  1204.2706}}].

\bibitem{Kitabayashi:2015jdj}
T.~Kitabayashi and M.~Yasuè, \emph{{Formulas for flavor neutrino masses and
  their application to texture two zeros}},
  \href{http://dx.doi.org/10.1103/PhysRevD.93.053012}{\emph{Phys. Rev.} {\bf
  D93} (2016) 053012}, [\href{http://arxiv.org/abs/1512.00913}{{\tt
  1512.00913}}].

\bibitem{Zhou:2015qua}
S.~Zhou, \emph{{Update on two-zero textures of the Majorana neutrino mass
  matrix in light of recent T2K, Super-Kamiokande and NO$\nu$A results}},
  \href{http://dx.doi.org/10.1088/1674-1137/40/3/033102}{\emph{Chin. Phys.}
  {\bf C40} (2016) 033102}, [\href{http://arxiv.org/abs/1509.05300}{{\tt
  1509.05300}}].

\bibitem{Singh:2016qcf}
M.~Singh, G.~Ahuja and M.~Gupta, \emph{{Revisiting the texture zero neutrino
  mass matrices}}, \href{http://dx.doi.org/10.1093/ptep/ptw180}{\emph{PTEP}
  {\bf 2016} (2016) 123B08}, [\href{http://arxiv.org/abs/1603.08083}{{\tt
  1603.08083}}].

\bibitem{Esteban:2016qun}
I.~Esteban, M.~C. Gonzalez-Garcia, M.~Maltoni, I.~Martinez-Soler and
  T.~Schwetz, \emph{{Updated fit to three neutrino mixing: exploring the
  accelerator-reactor complementarity}},
  \href{http://dx.doi.org/10.1007/JHEP01(2017)087}{\emph{JHEP} {\bf 01} (2017)
  087}, [\href{http://arxiv.org/abs/1611.01514}{{\tt 1611.01514}}].

\bibitem{nufitweb}
I.~Esteban, M.~C. Gonzalez-Garcia, M.~Maltoni, I.~Martinez-Soler and
  T.~Schwetz, \emph{{NuFIT 3.2 (2018), www.nu-fit.org.}}, .

\bibitem{deSalas:2017kay}
P.~F. de~Salas, D.~V. Forero, C.~A. Ternes, M.~Tortola and J.~W.~F. Valle,
  \emph{{Status of neutrino oscillations 2018: first hint for normal mass
  ordering and improved CP sensitivity}},
  \href{http://arxiv.org/abs/1708.01186}{{\tt 1708.01186}}.

\bibitem{Capozzi:2018ubv}
F.~Capozzi, E.~Lisi, A.~Marrone and A.~Palazzo, \emph{{Current unknowns in the
  three neutrino framework}},  \href{http://arxiv.org/abs/1804.09678}{{\tt
  1804.09678}}.

\bibitem{Feroz:2008xx}
F.~Feroz, M.~P. Hobson and M.~Bridges, \emph{{MultiNest: an efficient and
  robust Bayesian inference tool for cosmology and particle physics}},
  \href{http://dx.doi.org/10.1111/j.1365-2966.2009.14548.x}{\emph{Mon. Not.
  Roy. Astron. Soc.} {\bf 398} (2009) 1601--1614},
  [\href{http://arxiv.org/abs/0809.3437}{{\tt 0809.3437}}].

\bibitem{Feroz:2013hea}
F.~Feroz, M.~P. Hobson, E.~Cameron and A.~N. Pettitt, \emph{{Importance Nested
  Sampling and the MultiNest Algorithm}},
  \href{http://arxiv.org/abs/1306.2144}{{\tt 1306.2144}}.

\bibitem{Hagedorn:2004ba}
C.~Hagedorn, J.~Kersten and M.~Lindner, \emph{{Stability of texture zeros under
  radiative corrections in see-saw models}},
  \href{http://dx.doi.org/10.1016/j.physletb.2004.06.094}{\emph{Phys. Lett.}
  {\bf B597} (2004) 63--72}, [\href{http://arxiv.org/abs/hep-ph/0406103}{{\tt
  hep-ph/0406103}}].

\bibitem{Krolikowski:1999cx}
W.~Krolikowski, \emph{{Fermion texture and sterile neutrinos}}, {\emph{Acta
  Phys. Polon.} {\bf B30} (1999) 2631--2669},
  [\href{http://arxiv.org/abs/hep-ph/9903209}{{\tt hep-ph/9903209}}].

\bibitem{Zhang:2013mb}
Y.~Zhang, \emph{{Majorana neutrino mass matrices with three texture zeros and
  the sterile neutrino}},
  \href{http://dx.doi.org/10.1103/PhysRevD.87.053020}{\emph{Phys. Rev.} {\bf
  D87} (2013) 053020}, [\href{http://arxiv.org/abs/1301.7302}{{\tt
  1301.7302}}].

\bibitem{Nath:2015emg}
N.~Nath, M.~Ghosh and S.~Gupta, \emph{{Understanding the masses and mixings of
  one-zero textures in 3 + 1 scenario}},
  \href{http://dx.doi.org/10.1142/S0217751X16501323}{\emph{Int. J. Mod. Phys.}
  {\bf A31} (2016) 1650132}, [\href{http://arxiv.org/abs/1512.00635}{{\tt
  1512.00635}}].

\bibitem{Lavoura:2004tu}
L.~Lavoura, \emph{{Zeros of the inverted neutrino mass matrix}},
  \href{http://dx.doi.org/10.1016/j.physletb.2005.01.047}{\emph{Phys. Lett.}
  {\bf B609} (2005) 317--322}, [\href{http://arxiv.org/abs/hep-ph/0411232}{{\tt
  hep-ph/0411232}}].

\bibitem{Branco:2002ie}
G.~C. Branco, R.~Gonzalez~Felipe, F.~R. Joaquim and T.~Yanagida,
  \emph{{Removing ambiguities in the neutrino mass matrix}},
  \href{http://dx.doi.org/10.1016/S0370-2693(03)00572-0}{\emph{Phys. Lett.}
  {\bf B562} (2003) 265--272}, [\href{http://arxiv.org/abs/hep-ph/0212341}{{\tt
  hep-ph/0212341}}].

\bibitem{He:2003nt}
X.-G. He and A.~Zee, \emph{{Neutrino masses with 'zero sum' condition: m(nu(1))
  + m(nu(2)) + m(nu(3)) = 0}},
  \href{http://dx.doi.org/10.1103/PhysRevD.68.037302}{\emph{Phys. Rev.} {\bf
  D68} (2003) 037302}, [\href{http://arxiv.org/abs/hep-ph/0302201}{{\tt
  hep-ph/0302201}}].

\bibitem{Kaneko:2005yz}
S.~Kaneko, H.~Sawanaka and M.~Tanimoto, \emph{{Hybrid textures of neutrinos}},
  \href{http://dx.doi.org/10.1088/1126-6708/2005/08/073}{\emph{JHEP} {\bf 08}
  (2005) 073}, [\href{http://arxiv.org/abs/hep-ph/0504074}{{\tt
  hep-ph/0504074}}].

\bibitem{Chauhan:2006uf}
B.~C. Chauhan, J.~Pulido and M.~Picariello, \emph{{Neutrino mass matrices with
  vanishing determinant}},
  \href{http://dx.doi.org/10.1103/PhysRevD.73.053003}{\emph{Phys. Rev.} {\bf
  D73} (2006) 053003}, [\href{http://arxiv.org/abs/hep-ph/0602084}{{\tt
  hep-ph/0602084}}].

\bibitem{Lashin:2007dm}
E.~I. Lashin and N.~Chamoun, \emph{{Zero minors of the neutrino mass matrix}},
  \href{http://dx.doi.org/10.1103/PhysRevD.78.073002}{\emph{Phys. Rev.} {\bf
  D78} (2008) 073002}, [\href{http://arxiv.org/abs/0708.2423}{{\tt
  0708.2423}}].

\bibitem{Han:2017wnk}
J.~Han, R.~Wang, W.~Wang and X.-N. Wei, \emph{{Neutrino mass matrices with one
  texture equality and one vanishing neutrino mass}},
  \href{http://dx.doi.org/10.1103/PhysRevD.96.075043}{\emph{Phys. Rev.} {\bf
  D96} (2017) 075043}, [\href{http://arxiv.org/abs/1705.05725}{{\tt
  1705.05725}}].

\bibitem{Singh:2016cbe}
M.~Singh and R.~R. Gautam, \emph{{Exploring Texture Two-Zero Majorana Neutrino
  Mass Matrices with the Latest Neutrino Oscillation Data}},
  \href{http://dx.doi.org/10.1007/978-3-319-25619-1_50}{\emph{Springer Proc.
  Phys.} {\bf 174} (2016) 323--327}.

\bibitem{Bilenky:1999wz}
S.~M. Bilenky, C.~Giunti, W.~Grimus, B.~Kayser and S.~T. Petcov,
  \emph{{Constraints from neutrino oscillation experiments on the effective
  Majorana mass in neutrinoless double beta decay}},
  \href{http://dx.doi.org/10.1016/S0370-2693(99)00998-3}{\emph{Phys. Lett.}
  {\bf B465} (1999) 193--202}, [\href{http://arxiv.org/abs/hep-ph/9907234}{{\tt
  hep-ph/9907234}}].

\bibitem{Vissani:1999tu}
F.~Vissani, \emph{{Signal of neutrinoless double beta decay, neutrino spectrum
  and oscillation scenarios}},
  \href{http://dx.doi.org/10.1088/1126-6708/1999/06/022}{\emph{JHEP} {\bf 06}
  (1999) 022}, [\href{http://arxiv.org/abs/hep-ph/9906525}{{\tt
  hep-ph/9906525}}].

\bibitem{Vagnozzi:2017ovm}
S.~Vagnozzi, E.~Giusarma, O.~Mena, K.~Freese, M.~Gerbino, S.~Ho et~al.,
  \emph{{Unveiling $\nu$ secrets with cosmological data: neutrino masses and
  mass hierarchy}},
  \href{http://dx.doi.org/10.1103/PhysRevD.96.123503}{\emph{Phys. Rev.} {\bf
  D96} (2017) 123503}, [\href{http://arxiv.org/abs/1701.08172}{{\tt
  1701.08172}}].

\bibitem{Grimus:2004az}
W.~Grimus and L.~Lavoura, \emph{{On a model with two zeros in the neutrino mass
  matrix}}, \href{http://dx.doi.org/10.1088/0954-3899/31/7/014}{\emph{J. Phys.}
  {\bf G31} (2005) 693--702}, [\href{http://arxiv.org/abs/hep-ph/0412283}{{\tt
  hep-ph/0412283}}].

\bibitem{Xing:2002ta}
Z.-z. Xing, \emph{{Texture zeros and Majorana phases of the neutrino mass
  matrix}}, \href{http://dx.doi.org/10.1016/S0370-2693(02)01354-0}{\emph{Phys.
  Lett.} {\bf B530} (2002) 159--166},
  [\href{http://arxiv.org/abs/hep-ph/0201151}{{\tt hep-ph/0201151}}].

\end{thebibliography}\endgroup

\end{document}